%% file: main.tex
\documentclass[conference]{IEEEtran}
\pdfoutput=1
\IEEEoverridecommandlockouts
\usepackage{cite}
\usepackage{amsmath,amssymb,amsfonts}
\usepackage{graphicx}
\usepackage{textcomp}
\usepackage{multirow}
\usepackage{booktabs}
\usepackage{adjustbox}
\usepackage{dirtytalk}
\usepackage{float}
\usepackage{makecell}
\usepackage{enumitem}
\usepackage{svg}
\usepackage[ddmmyy]{datetime}
\usepackage{soul}
\usepackage{algorithm}
\usepackage[noend]{algpseudocode}
\usepackage{ragged2e}
\usepackage{setspace}
\usepackage{xcolor}
\usepackage{hyperref}

\newcommand{\lmttfont}{\fontfamily{lmtt}\selectfont}

\usepackage{caption}
\usepackage{subcaption}
\usepackage{xcolor}
\hypersetup{
    colorlinks,
    linkcolor={red!50!black},
    citecolor={blue!50!black},
    urlcolor={blue!80!black}
}

\def\BibTeX{{\rm B\kern-.05em{\sc i\kern-.025em b}\kern-.08em
    T\kern-.1667em\lower.7ex\hbox{E}\kern-.125emX}}

\def\sectionautorefname{Section}

\begin{document}



\title{Introducing k4.0s: a Model for Mixed-Criticality Container Orchestration in Industry 4.0}


\author{
\IEEEauthorblockN{Marco Barletta, Marcello Cinque, Luigi De Simone, Raffaele Della Corte}
\IEEEauthorblockA{\textit{Universit\`a degli Studi di Napoli Federico II, Italy} \\ %
\{marco.barletta, macinque, luigi.desimone, raffaele.dellacorte2\}@unina.it}
}

\newcommand{\mb}[1]{{\textcolor{green} {MB says: \textbf{#1}}}}
\newcommand{\gigi}[1]{{\textcolor{blue} {Gigi says: \textbf{#1}}}}
\newcommand{\raf}[1]{{\textcolor{red} {Raf says: \textbf{#1}}}}

\maketitle

\begin{abstract}
Time predictable edge cloud is seen as the answer for many arising needs in Industry 4.0 environments, since it is able to provide flexible, modular, and reconfigurable services with low latency and reduced costs.
Orchestration systems are becoming the core component of clouds since they 
take decisions on the placement and lifecycle of software components. Current solutions start introducing real-time containers support for time predictability; however, these approaches lack of determinism as well as support for workloads requiring multiple levels of assurance/criticality. 

In this paper, we present k4.0s, an orchestration model for real-time and mixed-criticality environments, which includes timeliness, criticality and network requirements. The model leverages new abstractions for both node and jobs, e.g., node assurance, 
and requires novel monitoring strategies. 
We sketch an implementation of the proposal based on Kubernetes, and present an experimentation motivating the need for node assurance levels and adequate monitoring. 
\end{abstract}

\section{Introduction}
\label{sec:intro}
\input{1_introduction}

\section{Problem Statement}
\label{sec:problem}
\input{2_problem_statement}

\section{System Model}
\label{sec:model}
\input{3_model}

\section{Proposed Architecture}
\label{sec:architecture}
\input{4_architecture}

\subsection{Implementation Details}
\label{subsec:implementation}
\input{5_implementation}

\section{Experimental Results}
\label{sec:results}
\input{6_results}

\section{Related Work}
\label{sec:related}
\input{7_related}

\section{Conclusion}
\label{sec:conclusion}
\input{8_conclusion}

\input{9_trash}


\bibliographystyle{IEEEtran}
\bibliography{main}

\end{document}

%% file: 1_introduction.tex
In recent years, we are witnessing the increasing adoption of IT technologies in several industrial domains (e.g., railways, avionic, automotive), and together with recent trends such as Industrial Internet of Things (IIoT) and Fog/Edge computing, they are paving the way for the rising revolution of \textit{Industry 4.0} (I4.0) \cite{naha2018fog, iiot_survey, stavdas2022networked}. The I4.0 includes several scenarios ranging from high-available distributed control systems \cite{johansson22kubernetes}, industrial automation with multiple networked components distributed across multiple and heterogeneous devices \cite{eidenbenz2020latency}, digital twins-based production processes \cite{borghesi2021iotwins}. With this paradigm shift, fog/edge computing systems can be seen as mixed-criticality systems\cite{burns2022mixed} since they need to consolidate different functionalities on the same edge node, e.g., real-time control, machine learning algorithms, high-speed data acquisition, and so on, each of them with different needs and requirements \cite{cilardo2021virtualization}.



Lightweight virtualization solutions, such as containers, are gaining the limelight for consolidating components onto one computing platform since they promise to deliver low-latency, bandwidth-efficient, and resilient services with reduced overhead and higher scalability compared to classical hypervisor-based solutions \cite{cinque2021virtualizing, morabito2018consolidate,struhar2020rtcontainerssurvey,abeni2019container, cinque2019rt, cinque2021preventing, goldschmidt2018container, tasci2018container, barletta2022achieving}. Moreover, containers are seamlessly integrated into orchestration platforms, fostering reconfigurability (e.g., Docker Swarm \cite{docker_swarm}, Kubernetes \cite{k8s}). On the other hand, OS-level virtualization is known to suffer from a reduced degree of isolation compared to hypervisor-based virtualization; thus it cannot be seen as a panacea \cite{randal2020ideal}.

\textit{Orchestration systems} are paramount since they automatically place, deploy, monitor, and migrate the packaged software across the infrastructure, behaving as cloud operating systems and controlling the whole system \cite{casalicchio2019container, khan2017key, rodriguez2019orchestration, tiburski2021lightweight}. I4.0 requirements lead to the need for \textit{criticality-aware} orchestration solutions, able to cope with a highly distributed ecosystem while providing the highest isolation guarantees to critical tasks, preventing severe faults/attacks within low-critical domains to interfere against high-critical ones, through the joint use of different virtualization technologies. 

Recently, several studies leveraged Kubernetes for real-time cloud computing orchestration \cite{fiori2022kube, johansson22kubernetes, toka2021ultra, struhar21react, eidenbenz2020latency} since it is a fairly popular tool for managing containerized applications due to its rich feature set. 
Despite the existing literature introducing the support for deploying real-time containers, orchestration for time predictable cloud is still at the dawn. Specifically, current solutions lack determinism (e.g., time-sensitive placement and migration of resources), support for heterogeneous devices and their peculiarities (e.g., awareness of accelerators in deploy decisions), and capability of meeting the different degrees of isolation and assurance required by the workloads. Further, since IIoT is made of heterogeneous devices including Multiprocessor System-on-Chip (MPSoC), FPGAs and GPUs, with rich I/O capabilities, an orchestrator should cope with hybrid solutions that leverage different virtualization approaches in one product \cite{runx, katacontainers, cilardo2021virtualization}.

On this basis, we propose a novel model for real-time mixed-criticality container orchestration in fog/edge systems, called \textit{k4.0s}, taking full advantage of heterogeneous virtualization technologies like OS-level virtualization and hypervisor-based containers to meet specified requirements. Specifically, our contributions fill the gap in the literature by providing: 
\begin{itemize}
    \item An \textit{orchestration model} for fog/edge cloud infrastructures that focuses on different assurance levels, time-sensitive networks, and mixed-criticality real-time containers;
    \item A \textit{design architecture} and a \textit{Kubernetes-based implementation} of the proposed orchestration model;
    \item An \textit{analysis of edge nodes behaviors under different stressful conditions}, motivating the need for different assurance and criticality levels;
    \item \textit{Monitoring metrics selection} for accurately defining and supporting the orchestration strategies. 
\end{itemize}

\noindent
The rest of the paper is organized as follows. \sectionautorefname{}~\ref{sec:problem} introduces our problem statement. \sectionautorefname{}~\ref{sec:model} provides the orchestration model for mixed-criticality I4.0 infrastructure. 
\sectionautorefname{}~\ref{sec:architecture} describes the proposed architecture and the details of our Kubernetes-based implementation. \sectionautorefname{}~\ref{sec:results} provides an experimental analysis motivating our proposal. 
\sectionautorefname{}~\ref{sec:related}  discusses related work, while \sectionautorefname{}~\ref{sec:conclusion} concludes the paper.

%% file: 2_problem_statement.tex
In the context of fog/edge computing systems and IIoT in I4.0, container orchestration systems need to meet new requirements. Some examples are the support for heterogeneous nodes/devices, the ability to meet the different degrees of assurance required by the workloads, etc. Both containers and container orchestration technologies have been designed considering a different target; this leads to an inherent lack of abstractions that prevents the actual use of these technologies in real I4.0 environments. Considering the existing container orchestration solutions, we identify four main challenges:

\begin{itemize}[leftmargin=4mm]
    \item \textbf{Proper abstractions for worker nodes}: Worker nodes, i.e., the nodes that actually carry out assigned tasks, are usually considered equivalent, except for their resource configuration (e.g., available CPUs, memory). This abstraction is not appropriate in the context of I4.0 systems, where worker nodes can be drastically different, with different guarantees and execution times. For example, edge nodes running closer to sensors devices can guarantee timing constraints on data transfer, while guarantees on latencies may depend on the hardware/software configuration (e.g., presence of accelerators or real-time OS).  

    \item \textbf{Adequate monitoring strategies.} Given the need for new abstractions characterizing the worker nodes to face I4.0 systems requirements, the monitoring strategies should be improved with adequate metrics. Current solutions are mainly designed for best-effort workloads.

    \item \textbf{Network awareness.} Orchestration systems have limited awareness of networks, and usually deal with a single network of nodes. However, I4.0 environments may encompass several networks, each one with very different features, such as Time Sensitive Networks (TSN), Controller Area Network (CAN) bus, wireless channels, and so on. Therefore, the orchestration services should be able to adapt placement strategies depending on networks characteristics coupled with application needs.

    \item \textbf{Real-time control plane.} \textit{Control planes}, i.e., the part of an orchestration system that takes global decisions, are often designed to be scalable, but it does not provide any latency guarantee. 
    Thus, it should be enhanced to provide orchestration decisions in a timely and predictable way. 
\end{itemize}

\noindent
In this work, we focus on the first three issues in the proposed model, sketching a prototype implementation for the first two. 
To this aim, we provide a novel orchestration model that includes (i) a worker node abstraction suitable for fog/edge cloud infrastructure encompassing the assurance level of the node, (ii) monitoring strategies and metrics supporting the orchestration model, (iii) the support for time-sensitive networks, as well as (iv) the support for mixed-criticality real-time containers. 

%% file: 3_model.tex
In this section, we provide an abstract description of our system model, adopting the terminology used in \cite{rodriguez2019orchestration} for our reference architecture of an orchestration system. We assume a fog/edge cloud infrastructure (e.g., an I4.0 environment) that includes a cluster of compute resources, made up of \textit{worker nodes}, interconnected with one or multiple types of \textit{networks}. A set of \textit{master nodes}, i.e., cluster manager master in \cite{rodriguez2019orchestration}, is in charge of both collecting \textit{jobs} submitted by the users and assigning them to the right worker nodes, according to the job requirements and the available \textit{resources} of the worker nodes.

\noindent
$\blacksquare$ \textbf{Worker Nodes}. Our cluster is made up of $I$ worker nodes. Each node $WN_i \; i\in [0,I[$ is characterized by an \textit{assurance level} $A_i$, and a \textit{real-time capability} $RT_i\in\{RT,\text{non-}RT\}$. The assurance level is the degree of dependability that a worker node can provide based on its hardware/software characteristics. It can be decomposed into $A_i = \alpha_i + \beta_i + \gamma_i(t)$ where $\alpha_i$ and $\beta_i$ are assurance levels provided respectively by the hardware and the software, while $\gamma_i(t)$ is a time function that represents the assurance level given by the state of the system at a glance. $\gamma_i(t)$ allows the chance of temporarily reducing the node assurance when a co-located workload turns out to potentially undermine the guarantees of a higher critical workload, and thus to represent a hazardous condition. The \textit{real-time capability} for a worker node specifies whether it is equipped with hardware/software that caters temporal guarantees.

\noindent
$\blacksquare$ \textbf{Resources}. Each worker node $WN_i$ is also characterized by a vector of \textit{basic hardware resources}, namely CPU, disk and memory, $\vec{BR}_i=[CPU, Disk, Mem]$ and a set of \textit{additional heterogeneous resources} ${AR}_i=\{\vec{S}, \vec{A}, GPU, FPGA\}\; i\in [0,I[$, where $\vec{S}$ and $\vec{A}$ are respectively vectors of sensors and actuators usually involved in an I4.0 environment. 
$\vec{BR}$ includes values representing the absolute available quantity of the resource. For CPU, we use the \textit{CPU utilization}, defined as $U=C/T$, where $C$ is the computation time and $T$ is the period. Depending on the scheduling algorithms used, \textit{CPU utilization} could be divided by the \textit{number of CPU cores}.
$AR_i$ is made up of integer values representing how many free resources are available in the node. We assume an exclusive allocation of a resource to a \textit{job} (definition ahead) for the sake of simplicity. If a resource can be shared between at most $K$ jobs, it could be simply modeled as $K$ separate resources.

\noindent
$\blacksquare$ \textbf{Networks}. Worker nodes are connected by one to multiple $Q$ networks $NW_q,  \; q \in [0,Q[$, with different purposes, requirements, and features. We can consider a classical \textit{best-effort network} connected with the Internet towards the centralized cloud, but also other networks with real-time guarantees, such as a TSN, a CAN bus, or a monitoring network. $NW_q$ is a graph described by $(V_q, E_q)$. $V_q = \{WN_0, ..., WN_{|V_q|}\}$ is the set of nodes belonging to the network, and $E_q$ are the links between nodes. Links are characterized by a communication delay $E^d_q$ and a capacity $E^c_q$, defining, in general, a matrix $M_q:V_q^2\rightarrow D$, where D is the communication delay.
In our model, we advocate that networks should be treated distinctively by the orchestration systems because of their different requirements. For example, in a real-time network, a sound schedulability analysis is needed to enforce temporal guarantees, as it happens in TDMA networks, typical of various real-time networking standards \cite{ergen2010tdma}. Also, for a best-effort network, there could be bandwidth or delay constraints for a job. Thus, we are not tight to a specific network model for a specific network type.

\noindent
$\blacksquare$ \textbf{Jobs}. On each worker node $WN_i$, a set of $J_i$ jobs is scheduled. A \textit{job} $J_{ij} \; i \in [0,I[,\; j \in [0,J_i[$ consists of a set of containers with different running tasks (defined ahead) into. A job is the \textit{minimum unit of scheduling for the orchestrator}, and it has a set of allotted resources. We treat a job without caring about the internal composition. 
A job relies on allocated basic resources $\vec{BR}_{ij}$ to complete its activities, and it can (optionally) take advantage of its additional resources $AR_{ij}$. While it is trivial to define the memory or disk needed by a job as a sum of resources needed by its tasks, CPU requirements need a more detailed analysis, later discussed. Finally, a job could have a set of specific communication requirements $NW^{req}=\{NW^{req}_0, ...,NW^{req}_q\}$ on various networks. Specifying how the requirements should be expressed is strictly related to the network type.

A job can optionally have real-time requirements ($RT_{ij}=RT$). \textit{Real-time jobs} can be placed only on real-time--enabled worker nodes. We further introduce the notion of \textit{job criticality}: different replicas of a job can be deployed, each of them having a different criticality levels $C_{ij} \in \{NO, LOW, HI\}$. We assume that real-time jobs have at least the $LOW$ criticality. A set of replicas, namely a \textit{deployment}, could be characterized by a \textit{leader replica} with a higher criticality level, while other replicas, with lower criticalities, can serve as backups, or implement Triple Modular Redundancy (TMR). Each criticality level should be backed by an appropriate assurance level of the worker node: a high criticality job should be deployed on a node with a high assurance $A_i$ and should be implemented through a virtualization layer that provides a high degree of isolation, e.g., hypervisor-based containers; less critical jobs could take advantage of dynamic environments, with fewer guarantees.
Finally, a job is described by a tuple $J_{ij}=<\vec{BR}, AR, NW_q, C_{ij}, RT_{ij}>$

\noindent
$\blacksquare$ \textbf{Tasks}. A job is made up of a set of containers, each containing different tasks $TK_{ijk}\; i \in [0,I[,\; j \in [0,J_i[,\; \forall k $. In the case of a job with timeliness requirements, its real-time tasks are described by an additional \textit{real-time scheduling interface}. Differently from previous studies, we believe that the real-time scheduling interface of a job can not be described by a model as simple as the $(period, budget)$ pair: the budget needed by a container heavily depends on the execution time, which changes drastically upon the hardware, especially in a heterogeneous environment such as I4.0. Moreover, how the CPU bandwidth of tasks is combined to obtain the total CPU bandwidth needed by the whole job depends on the scheduling algorithm used. To overcome these limitations, in the proposed model each task $TK$ is described by a period $T$ and a vector of worst-case execution times (WCET). Similarly to the approach of the Vestal model (\cite{vestal2008}), the execution times depend on the criticality level: for high-criticality workloads, the WCET should be precisely computed for a set of specific hardware platforms, and only these platforms should be taken into account during the scheduling phase. However, a precise estimation of execution times on every platform is infeasible, thus a low-criticality workload could benefit from a less precise but less expensive method for WCET estimation: an option would be rescaling a baseline WCET by a factor that depends on the platform. In this way, only a factor common to multiple workloads should be derived, with much less effort. 
Anyway, defining the precise WCET computation methods is beyond the scope of this work, and we generally assume a WCET that is a function of both the criticality level and the worker node considered, $WCET_{ijk}(i,j)$. 

The task requirements for both $\vec{BR}$ and $AR$ are summed up and seen as requests for the whole job. Hence, we model a task as a tuple $TK_{ijk}=<WCET_{ijk}(i,j), T_{ijk}, P_{ijk}>$, where $T$ is the period, $P$ the (optional) task priority. Assuming a scheduling interface for each task, we are completely independent of how the scheduling framework joins task parameters to obtain a CPU request for the whole job. In this way, we encompass the use of a plethora of \textit{schedulability tests} on the nodes, with the only requirement that tasks fit into the real-time scheduling interface, and master nodes are not even aware of the specific scheduling algorithm used on the worker nodes.

\begin{table}[ht!]
\centering
\begin{adjustbox}{max width=\columnwidth}
\begin{tabular}{cc} 
\textbf{Symbol} & Meaning\\ 
\toprule\toprule
$i\in [0,I[ \quad j\in [0,J_i[\quad q \in [0,Q[\quad \forall k$ & indexes of nodes, jobs, nets, tasks\\
\hline
$WN_i = <RT_{i},A_{i},\vec{BR}_i,{AR}_i>$ & worker node $i$\\
\hline
$RT_i\in\{RT,\text{non-}RT\}$ & real-time capability of node $i$\\
\hline
$A_i = \alpha_i + \beta_i +\gamma_i(t)$ & assurance of node $i$\\
\hline
$\vec{BR}_i=[CPU,Disk,Mem]$ & free basic resources of node $i$\\
\hline
${AR}_i=\{\vec{S},\vec{A},GPU,FPGA\}$ & free additional resources of node $i$\\
\hline
$NW_q = (V_q,E_q)$ & network $q$\\
\hline
$M_q:V_q^2\rightarrow D$ & delay matrix of network $q$ \\
\hline
$J_{ij}=<\vec{BR}, AR, NW^{req}, C_{ij},RT_{ij}>$ & job $j$ on node $i$\\
\hline
$C_{ij} \in \{NO, LOW, HI\}$ & criticality of job $j$ on node $i$\\
\hline
$TK_{ijk}=<WCET_{ijk}(i,j),T_{ijk}, P_{ijk}>$ & task $k$, job $j$ on node $i$\\
\hline
$WCET_{ijk}(i,j)$ & WCET of task $k$, job $j$ on node $i$\\ 
\hline
$T_{ijk}, P_{ijk}$ & period and priority of task $k$\\
\hline
\end{tabular}
\end{adjustbox}
\vspace{5pt}
\caption{Parameters of the system model.}
\label{tab:latency}
\vspace{-10pt}
\end{table}

 

%% file: 4_architecture.tex
An orchestration system can be seen as two logical components: a \textit{control plane} and a \textit{compute cluster}. The former is the part of the system that receives requests for deployment, monitors the compute cluster, and manages its state by coordinating the lifecycle of the jobs. Control plane is deployed onto one or more (for dependability concerns) consensual \textit{master nodes}. The compute cluster actually carries out the work assigned, and it is physically composed of the worker nodes on which jobs are scheduled.
\begin{figure}[!ht]
    \centering
    \includegraphics[scale=0.87]{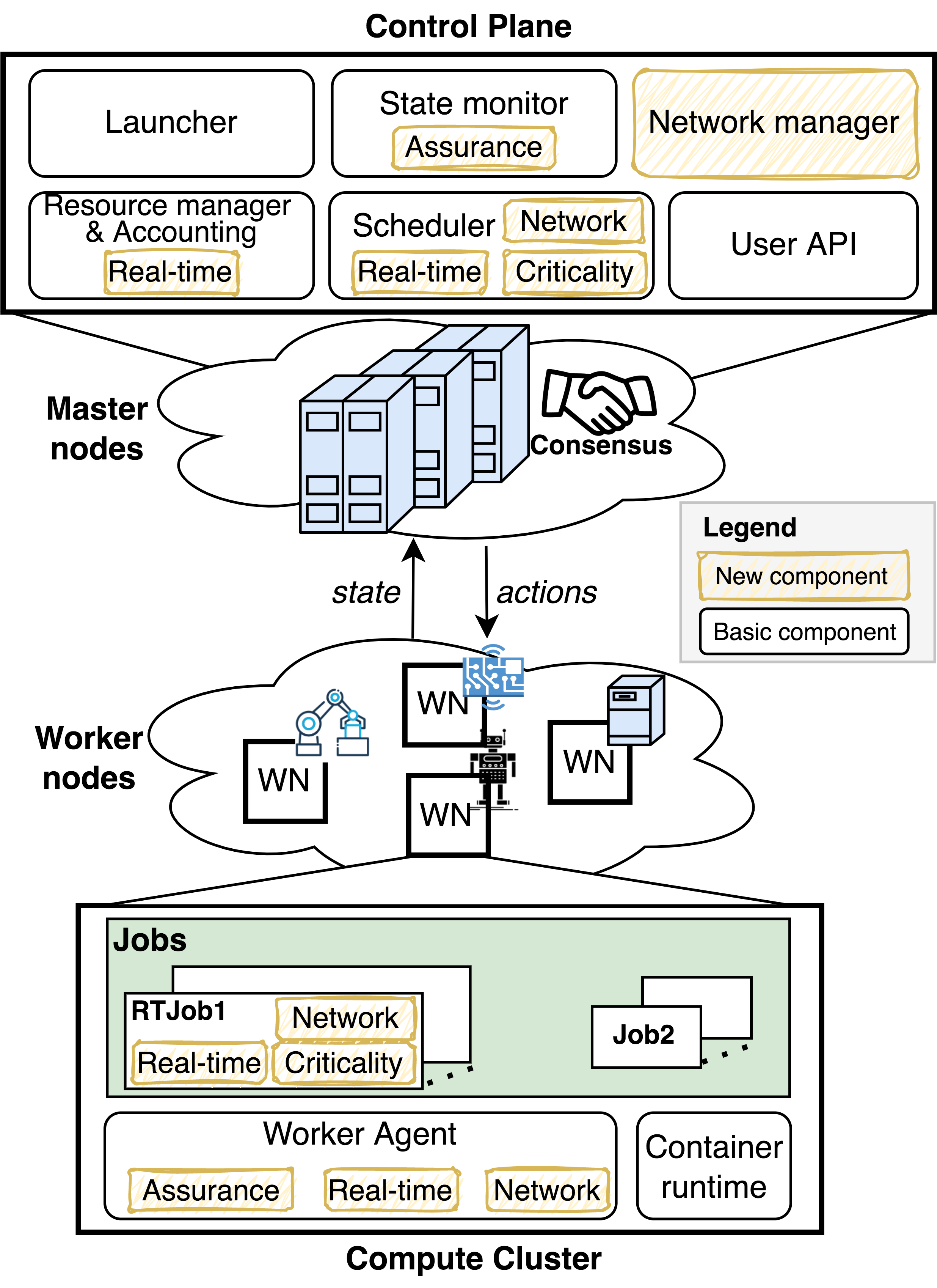}
    \caption{Architecture of k4.0s.}
    \label{fig:architecture}
    \vspace{-15pt}
\end{figure}
\figurename~\ref{fig:architecture} shows a simplified view of the architecture of an orchestration system from \cite{rodriguez2019orchestration}, with our modifications highlighted. The control plane alters the state of the compute cluster through the actions issued towards the worker nodes, while the latter periodically notifies the control plane about the state of both jobs deployed and worker nodes.

\subsection{Control Plane Architecture}

The control plane is composed of the following communicating components: \textit{i)} a \textit{Launcher} that issues decided actions to the worker nodes; \textit{ii)} a \textit{State monitor} that stores the global state of the worker nodes and jobs, and issues actions when the state deviates from the desired one; \textit{iii)} a \textit{Resource manager \& Accounting} that keeps track of compute cluster resources, being aware of available ones for both scheduling and accounting purposes; \textit{iv)} a \textit{Scheduler} that verifies if there are enough resources for a job and decides its placement. It is also involved in the rescheduling of jobs when a migration is needed; \textit{v)} a \textit{User API}, that receives deployment requests, validates them, and triggers the scheduler if needed.

This basic scheme must be integrated with a new set of components to support the abstractions introduced in \sectionautorefname~\ref{sec:model}. We argue that the following extensions are needed. 

\noindent
$\blacktriangleright$ \textbf{Criticality-aware scheduler}. The \textit{Scheduler} is extended to support different placement policies for each criticality level. This is needed since placement policies often rely on heuristics, 
and the correct strategy must depend on the criticality of a job. For example, for non-critical jobs consolidation may be more important than isolation, achieving a better resource utilization and lower energy consumption. Conversely, critical jobs must be scheduled on nodes with high assurance.
This could in turn lead towards a \textit{spread} or \textit{pack} strategy, minimizing either the utilization or the number of worker nodes involved. Further, the criticality level is paired with a preemption level, which allows critical workloads to preempt lower critical ones in case of lack of resources.
    
 \noindent   
$\blacktriangleright$ \textbf{Real-time scheduling support}. The \textit{Scheduler} and  \textit{Resource manager \& Accounting} need to support real-timeliness to meet temporal constraints. 
The scheduler detects the real-time scheduling interface of jobs and forwards requests to a dedicated entity, which computes the schedulability test. This entity is an \textit{Agent} placed either on master nodes or on each worker node, aware only of local information, with possibly intermediate solutions. We adopt the second strategy.
The scheduler is not aware of the particular schedulability test used on each worker node, fostering flexibility and scalability with \textit{pluggable} schedulability tests. This enables the use of a broad range of virtualization approaches, each characterized by different real-time frameworks. 

\noindent
$\blacktriangleright$ \textbf{Assurance monitor}. This module collects information about the assurance of the worker nodes and their real-time capability. This information is used by the \textit{Scheduler}, through some heuristics: highest assurance nodes could be first selected for high criticality workloads, while least assurance nodes satisfying requirements are preferred for low criticality ones.

\noindent
$\blacktriangleright$ \textbf{Network manager}. This component manages virtual networks and keeps track of the physical topology and the different types of networks. The scheduler delegates to it the schedulability test strictly related to the network requirements. 
For example, the \textit{Network manager} keeps track of timeslots used in a TDMA-based network, delay matrices for best-effort networks, or available capacity for each physical link. Usually, orchestration systems also provide network services such as load balancers and proxies to cope with complex virtualized networks. We introduce similar components, suitable for critical applications. In particular, we add a \textit{replica manager} that keeps track of the master and backup replicas of jobs, and forwards requests to the proper one, and a \textit{TMR voter} that compares job responses and decides on majority.

\subsection{Compute Cluster Architecture}
\label{subsec:compute_arch}

The compute cluster is made up of worker nodes running:\\ \textit{i)} a \textit{Worker agent}, which collects health information about jobs and worker nodes itself, and forwards it to the control plane. Further, it communicates with the container runtime to deploy jobs and keep the desired state; \textit{ii)} a \textit{Container runtime}, that is a container engine (e.g., Docker); \textit{iii)} the jobs deployed. We extend the architecture as described in the following.

\noindent
$\blacktriangleright$ \textbf{Assurance module}. In the worker agent, this module extends monitoring capabilities by sampling system-level metrics useful to unveil if the worker node is appropriate for real-time and critical workloads. With this extension, the monitor could also leverage application-level metrics for a more accurate analysis; to this aim, the job is in charge of gathering them. The tuning of monitoring parameters (e.g., sampling period) should be done accordingly to the isolation guarantees provided by the underlying system.
The monitoring process is split into three stages. First, anomalies are detected through periodic sampling and trend estimation, raising alarms 
if the state is deemed hazardous for a real-time workload. The alarm triggers the second stage, which tries to identify and kill the guilty lower critical workloads. If the monitor still judges the state as hazardous after a predefined timeout, it downgrades the assurance level of the worker node and notifies the control plane to migrate the critical workload.
    
\noindent
$\blacktriangleright$ \textbf{Real-time module}. In the worker agent, this module accepts real-time requests for jobs, computes the schedulability test, and reserves real-time resources. How to leverage real-time resources is strictly dependent on the virtualization technology. 

\noindent
$\blacktriangleright$ \textbf{Network-aware module}. This module should be already within the worker agent; however, it must be extended with the support for other classes of networks, such as TSN, TDMA and so on. Further, it is responsible for allotting network resources to jobs.

\noindent
$\blacktriangleright$ \textbf{Job abstraction}. The job is extended with criticality, real-time, and network requirements. Thus, a request for a job deployment contains \textit{i)} the criticality level, \textit{ii)} the real-time scheduling interface for each task, including (in the case of a high-critical job) the WCET for the solely worker nodes where it has been computed, 
and \textit{iii)} the network requirements, e.g., the length $TS_{ij}$ and the period $TS_{p}$ for TDMA timeslots.

%% file: 5_implementation.tex
We implemented a prototype of the architecture described above by using Kubernetes since it is a \textit{de facto} standard in the container orchestration due to the great extensibility out of the box. In particular, for the control plane, we provide three \textit{scheduler plugins} that implement the criticality-aware scheduler, real-time scheduling support, and a TDMA-enabling network manager. Both master nodes and worker nodes are Kubernetes \textit{Nodes}, and jobs are implemented via Kubernetes \textit{Pods} within Nodes.

The criticality-aware scheduler filters out worker nodes with an unsuitable assurance level, then it scores them in decreasing/increasing order, looking for the assurance in the Node \textit{annotations} \cite{annotations}. 
The real-time scheduler plugin looks in the Pod annotations for real-time requirements, sends them to worker agents running on worker nodes in the scheduling request, and filters them based on the response. The response contains the utilization for each core along with schedulability result and allocated bandwidth if present. Finally, the scheduler scores the worker nodes using the remaining utilization for the least loaded core. 
The TDMA-enabling network manager relies on Pod annotations for TDMA network requirements, looks for a suitable timeslot, and then forwards the slot parameters to a TDMA master to set up resources.

The prototype includes three Kubernetes daemons to be deployed on worker nodes named \textit{Malacoda}, \textit{Rubicante}, and \textit{Scarmiglione} \cite{alighieri2007divina}. \textit{Malacoda} implements the three stages assurance monitor by modifying the node assurance level stored in the node annotations. \textit{Rubicante} implements the real-time module within the worker agent: it receives scheduling requests, computes the schedulability test for a hierarchical deferrable server algorithm, and allocates resources. Finally, \textit{Scarmiglione} implements a TDMA master logic, i.e., it receives timeslots parameters and creates the slot on a Xenomai RTnet network; we choose Xenomai as a solution to implement real-time containers\cite{barletta2022achieving}. 
We underline that we do not modify Kubernetes core, allowing us to replace the fully-featured Kubernetes with other Kubernetes-compliant distributions, such as k3s or k0s \cite{fathoni2019performance}. 
The source code is available at \cite{k4.0s}.

%% file: 6_results.tex

\begin{figure*}[!ht]
    \centering
    \includegraphics[width=1.8\columnwidth]{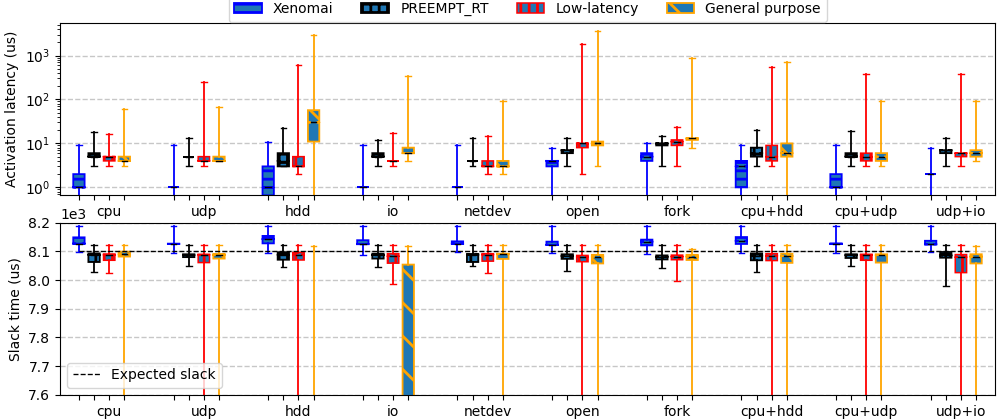}
    \caption{Task activation latencies (lower is better) and slack times (higher is better) under stress conditions for different systems. Slack times are clamped for readability.}
    \label{fig:latenciesandslacks}
    \vspace{-0.5cm}
\end{figure*}

In this section, we motivate the need for the assurance and criticality level in our model, as well as new monitoring strategies through experimental measures. We first show how real-time containers behave on a worker node configured with different assurance levels under stressful workloads, giving an insight into how practically the $\beta_i$ could be set. On this basis, we derive which workloads are deemed hazardous on different kernels and which metrics are suitable to represent the assurance level of the system. 
Our setup encompasses Intel Core i5-6500, 16 GB RAM, and Samsung 980 EVO SSD. 

\subsection{Isolation Comparison}

We run the \textit{rt-app} \cite{rtapp} for 30 min. to gather both the task activation latency and slack time, defined as $d-a-C$, where $d$ is the task deadline, $a$ is the arrival time, and $C$ its WCET. The run encompasses a single thread with a period of 10 ms and a runtime of 1.9 ms.
The 30 min. are split into 9 consecutive phases, each of them composed by 2 min. of cooldown and 1 min. of heavy co-located stress. In each phase, along \textit{rt-app}, we apply a different stress generated with \textit{stress-ng} \cite{stress} through the following options: \textit{cpu}, \textit{udp}, \textit{hdd}, \textit{io}, \textit{netdev}, \textit{open}, \textit{fork}, \textit{cpu} + \textit{hdd}, \textit{cpu} + \textit{udp}, \textit{udp} + \textit{io}. The last three workloads combine stresses belonging to the three categories IO, network and CPU, with the aim of detecting possible amplification or attenuation effects.
The experiment is repeated with an increasing number of stressor tasks: 1, 2, 4, 8, and 16.
We run the experiment 4 times on each target system: (i) Linux v5.4.77 with {\lmttfont rt-cgroups} \cite{rtcgroup} enabled, (ii) Linux v5.4.77 with {\lmttfont rt-cgroups} enabled and \textit{low-latency} preemption mode, (iii) Linux v5.15.32 with {\lmttfont rt-cgroups} enabled and {\lmttfont PREEMPT\_RT} patch, and (iv) Linux v5.4.77 patched with Xenomai v3.1. Besides the first configuration, all have power optimizations and frequency scaling disabled. 
\figurename~\ref{fig:latenciesandslacks} shows results for the maximum load (16 stressors).
Clearly, the general-purpose kernel can not provide any guarantee, and even worse, it behaves better under loads than in cooldown, probably due to frequency scaling. The \textit{low-latency} kernel behaves better, but there are still troublesome workloads that can be hazardous for a critical task. {\lmttfont PREEMPT\_RT} presents bounded errors, while Xenomai is the most predictable system, with very small interquartile ranges. 
In general, \textit{hdd} and \textit{open} are the most problematic workloads for all systems, while the effect of the others depends on the specific system. The results highlight that the considered systems provide increasing guarantees in our setup, which motivate the need for the node assurance level and $\beta_i$ value proposed in our model.

\subsection{Monitoring Metrics Selection}

\begin{figure}[!ht]
    \vspace{-0.5cm}
    \centering
    \includegraphics[width=0.8\columnwidth]{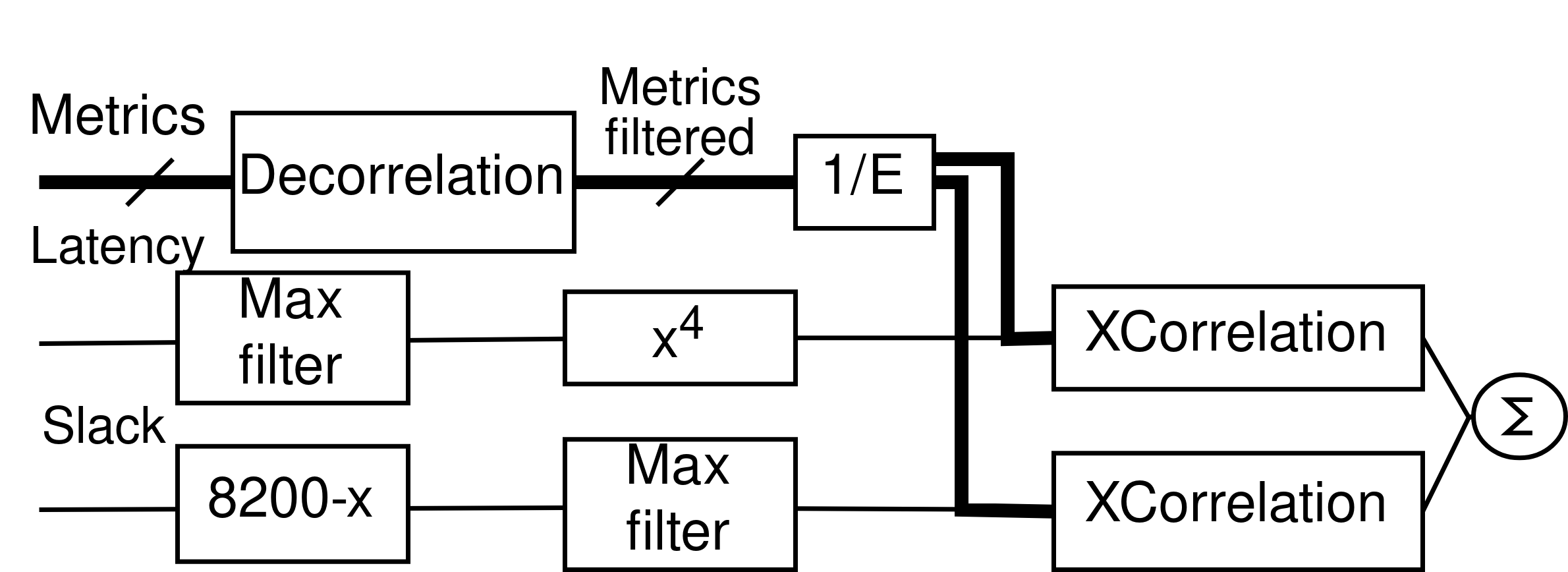}
    \caption{Block diagram for metric scoring.}
    \label{fig:metricblock}
    \vspace{-0.3cm}
\end{figure}

With data obtained from the previous test, we derived the most troublesome stresses for a real-time critical workload.
During experiments, we monitored each system, periodically sampling metrics from {\lmttfont /proc/stat}, {\lmttfont /proc/uptime}, {\lmttfont /proc/loadavg} and {\lmttfont /proc/softirqs} files.
We computed the cross-correlation between each monitoring metric (normalized to have unitary \textit{energy}, i.e., the sum of squares) and both the activation latency and the slack time as depicted in \figurename~\ref{fig:metricblock}. We filtered out highly correlated metrics from the process. 

\begin{figure}[!ht]
    \vspace{-5pt}
    \centering
    \includegraphics[width=0.9\columnwidth]{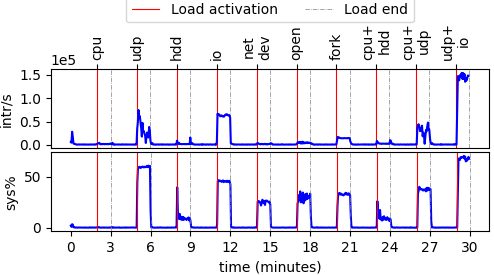}
    \caption{Examples of monitoring traces for \textit{sys\%} and \textit{intr/s}.}
    \label{fig:traces}
    \vspace{-20pt}
\end{figure}

The \textit{max filter} replaces each sample with the max in a sliding window of a determined size (50, in our case), and the exponentiation ($x^4$ block) has the aim to increase the weight of spikes in the traces. The final score used to rank the metrics for each system is computed as a weighted sum between the two correlations.
The metrics with the highest score can be used by the monitor to trigger alarms and detect the most troublesome workloads. For example, the top-5 metrics obtained in the \textit{low-latency} configuration are \textit{loadavg}, \textit{intr/s}, \textit{RES/s}, \textit{TIMER/s} and \textit{sys\%}; both \textit{sys\%}  and \textit{intr/s} measurements are shown in \figurename~\ref{fig:traces}. Since each target system provides a specific isolation degree and reacts variously under stress, the monitor should leverage a different set of metrics accordingly. 
Based on the key metrics, the monitor can temporarily allow stress conditions for a strongly isolated system; conversely, it must timely notify and react to any hazardous condition on a poorly isolated one, decreasing its $\gamma_i(t)$ when needed.


%% file: 7_related.tex
Many studies in the literature focus on time predictable cloud, trying to employ technologies such as containers and orchestration systems to implement and manage applications with delay critical constraints. The advantages are manifold and include scalability, flexibility, reconfigurability, increased reliability, reduction of costs and consumption of resources. 
In this context, different approaches have been proposed for the orchestration of real-time containers. 
For example, the work in \cite{fiori2022kube} presents \textit{RT-Kubernetes}, a modified version of the Kubernetes orchestrator allowing the deployment of real-time containers based on \textit{rt-cgroups}. RT-Kubernetes provides an extended description of tasks to include real-time constraints as well as an enhanced version of the Kubernetes scheduler and node agent, i.e., {\lmttfont kubelet}. Similarly, in \cite{struhar21react} Kubernetes has been modified to enable the deployment of both real-time and best-effort containers. To this aim, the Kubernetes control plane is modified to implement both admission control and scheduling of real-time containers, and an RT-manager is deployed on each working node to evaluate the node performance. This latter is performed using container-level and system-level metrics, which are analyzed during orchestration. The work in \cite{cucinotta21openstackffv} integrates real-time containers into a modified version of OpenStack, with the aim to foster the NFV management for next-generation networks. An architecture for real-time \textit{Function as a Service (FaaS)} is presented in \cite{szalay22rtfaas}, with the aim of opening cloud/edge computing infrastructures to time-critical applications. The proposal assumes functions running within containers; it leverages an analytical model that considers the end-to-end latency of real-time requests/responses pair, with particular stress on the importance of real-time communication. Furthermore, a partitioning algorithm is designed to minimize the number of compute nodes used. A Kubernetes-edge-scheduler is prototyped in \cite{toka2021ultra}. The scheduler enables reliability for edge applications through spare backup resources; it considers latency constraints of applications, extending the cluster scheduler with the awareness of a delay matrix. Similarly, in \cite{eidenbenz2020latency} is proposed a Kubernetes-based fog layer architecture for industrial automation contexts, which is latency-aware with regard to communication costs. The work aims to minimize the amount of data transferred between 
devices leveraging a greedy heuristic.

Differently from the existing real-time container orchestration systems, our proposal encompasses an orchestration model for fog/edge cloud infrastructures, which \textit{i)} extends the abstraction of worker nodes to include the assurance level, \textit{ii)} leverages a monitoring strategy enabling the collection of metrics to evaluate the assurance level of nodes and to drive the orchestration of real-time containers, \textit{iii)} contemplates different network types and requirements, and \textit{iv)} supports mixed-criticality containers.

%% file: 8_conclusion.tex
In this work, we introduced \textit{k4.0s}, a new orchestration model for mixed-criticality real-time containers, aimed at I4.0 scenarios. We reshaped the abstractions of an orchestration system to take into account the assurance of nodes, criticality of workloads, time-sensitive networks and a suitable real-time scheduling interface for the jobs. We extended the reference architecture for orchestration systems, introducing new components that deal with proposed abstractions, and sketched a Kubernetes-based implementation. We argued the need for new monitoring strategies, aware of the requirements of the workloads, and proposed a method to unveil key metrics, showing through measurements that different systems provide unequal isolation capabilities to containers. Future work will include the design of a real-time control plane by defining detailed monitoring and migration strategies.
